# Magnetization precession after non-collinear dual optical excitation

Sergii Parchenko[1, *], Davide Pecchio[2,3], Ritwik Mondal[4], Peter M. Oppeneer[5], Andreas Scherz[1]

1. *European XFEL, Holzkoppel 4, 22869 Schenefeld, Germany*
2. *Laboratory for Mesoscopic Systems, Department of Materials, ETH Zurich, 8093 Zurich, Switzerland*
3. *Laboratory for Multiscale Materials Experiments, Paul Scherrer Institute, 5232 Villigen PSI, Switzerland*
4. *Department of Physics, Indian Institute of Technology (ISM) Dhanbad, IN-826004, Dhanbad, India*
5. *Department of Physics and Astronomy, Uppsala University, Box 516, SE-75120 Uppsala, Sweden*

*Corresponding author: sergii.parchenko@xfel.eu

**Abstract. We investigate the impact of non-collinear dual optical excitation on the magnetization precession in a permalloy thin film using two ultrashort laser pulses. By analyzing the magnetization dynamics using time-resolved magneto-optical methods, we find that the excitation with two ultrashort optical pulses introduces a long-lasting modification of the electron system seen as a sizable decrease of the precession frequency and significant increase (about 25%) of the decay time. Our results reveal that the observed effect strongly depends on the respective polarization of the two excitation pulses and the time delay between two optical pulses. Our findings indicate the occurrence of a new nonlinear opto-spin effect during photoexcitation with two interfering optical pulses which can potentially be observed in various materials and at different photon wavelengths.**

The interaction of light with matter is one of the most fundamental processes in nature. From the absorption and emission of light by atoms and molecules to the propagation of light through materials, the properties of matter are intimately tied to the behavior of light and vice versa. Researchers have made significant progress in understanding light-matter interaction, particularly with the development of ultrafast lasers that generate light pulses with a duration on the order of femtoseconds, which makes it possible to study the dynamics of the interaction on a timescale that matches the response of the material[1]. However, one of the key challenges in studying and controlling light-matter interaction is the response of the material occurring on short timescales (femtoseconds to picoseconds).

Femtosecond laser pulses become an increasingly important tool to control material properties on ultrashort timescales. All-optical magnetization reversal [2,3,4], evidence of light-induced superconductivity [5,6], and optical-driven phase transitions [7, 8] are only some examples. While ultrashort pulses have been tuned over a wide range of wavelengths, allowing selective excitation of the electron and lattice degrees of freedom, a vast majority of studies use a single laser pulse to pump the system. Only very few studies explore the possibility of manipulating matter with two or more ultrashort pulses [9,10,11,12,13,14]. In such a two-step excitation process the first pump pulse promotes the material to an intermediate excited state, while the second pump pulse drives the system further out of equilibrium. In certain cases, this approach allows for a more efficient control of the material properties.

It is generally believed that the excitation with two optical pulses that overlap in time and space on the microscopic level behaves like a single-pulse excitation. In other words, the response of an electron when excited with two non-collinear propagating photons is expected to be the same as when it is excited with photons that propagate collinearly. Nonlinear optical effects such as the formation of transient gratings [15,16,17], difference, or sum frequency generation could take place, but the long-lasting effect of the excitation on the material is expected to follow simply the dependence determined by the strength of the electric or magnetic field of the incoming radiation.

Here, we demonstrate that optical excitation with two non-collinear light pulses can cause an unexpected response in a magnetic material. We show that the amplitude, decay time and frequency of magnetization precession in magnetically ordered alloys are changed when the dynamics is triggered by two non-collinear interfering optical pulses. The observations cannot be explained by the occurrence of a transient grating and indicate the existence of an unknown opto-spin effect that results in modified intrinsic magnetic properties of the material persisting much longer than the duration of the two coherent optical pulses.

We study the effect of dual non-collinear optical excitation on the widely explored physical process of laser-induced magnetization precession [18, 19]. This process involves the interaction between light, electrons, and eventually, spins, making the magnetization precession an ideal test bed for our considerations. The mechanism for triggering magnetization oscillations with an ultrashort laser pulse is based on the heat-induced reduction of the magnetization $M$ [20], resulting in a local change of the magnetic anisotropy and subsequent pointing of the magnetization away from the equilibrium position. This launches a magnetization precession that can be described by the Landau-Lifshitz-Gilbert equation and the excitation of spin waves [21]. We have extended this simple approach by implementing a second excitation pulse and varying the delay between the two optical pump pulses.

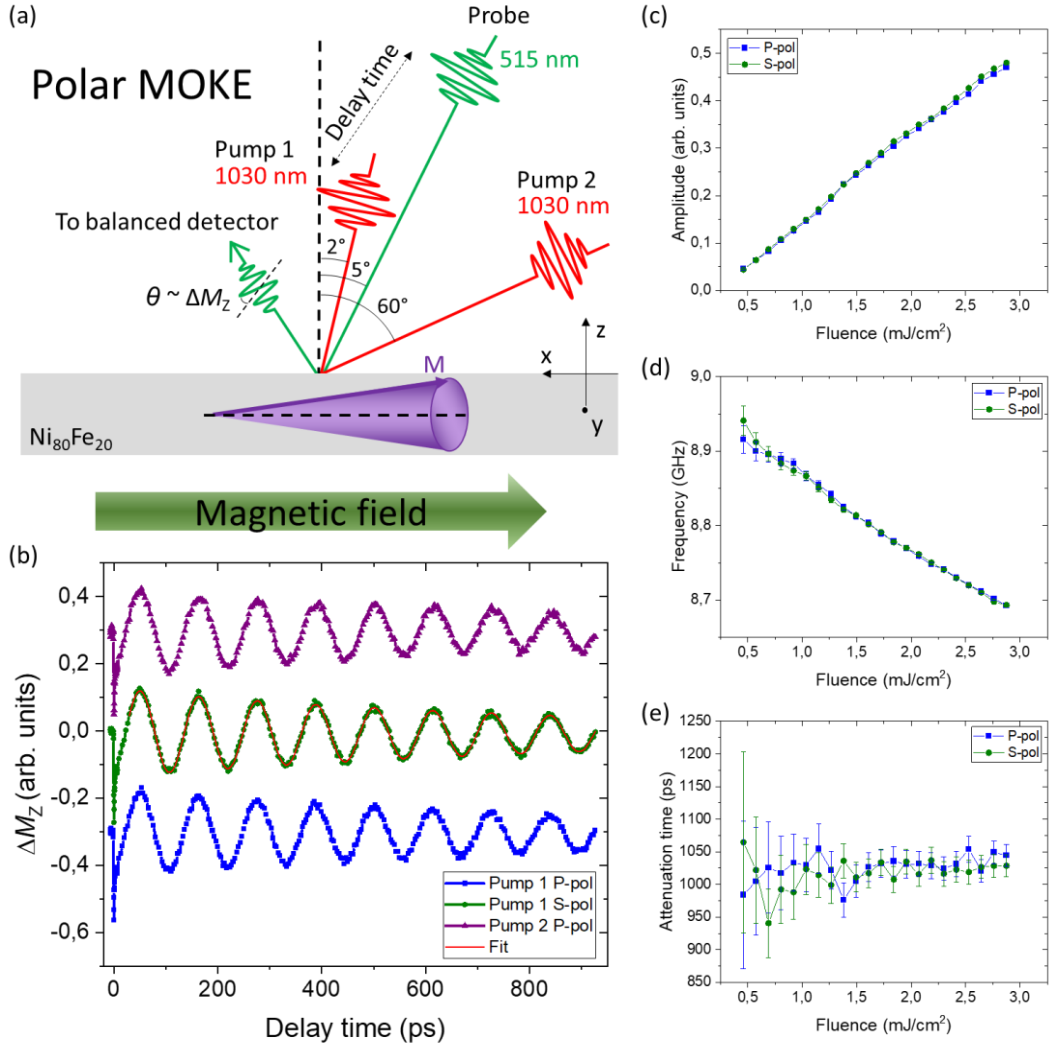

*Figure 1. (a) Schematic illustration of the experimental geometry. The permalloy sample is excited by two non-collinear laser pumps, and the magnetization dynamics is probed with the polar magneto-optical Kerr effect (P-MOKE). (b) Time-resolved magnetization dynamics after single pulse excitation for different pump polarization with an incident fluence of F= 0.92 mJ/cm² for pump 1 and F=1.5 mJ/cm² for pump 2. While the oscillation curves are the same in frequency and amplitude, indicating the equivalence of optical excitation with both pumps, the higher fluence for pump 2 is required due to different absorption at a 60° incidence angle. The curves are vertically shifted for better visibility. The red solid line is fit to eq. 1. (c) Amplitude, (d) frequency and (e) decay time of magnetization precession as a function of pump 1 fluence for different polarization.*

The material studied in our experiments is a permalloy (Py) thin film deposited on top of a $SiO_2$ substrate. Py, a $Ni_{80}Fe_{20}$ alloy with a thickness of approximately 20 nm, was prepared using a thermal evaporation method at a base pressure of $1\times10^{-6}$ mbar and was capped with a 2-nm-thick layer of gold to prevent oxidation. Py has aroused considerable interest due to observed magnetic monopole-like behavior in nanoscale patterned Py structures (Py artificial spin ice) [22,23,24], optical excitation of propagating spin waves [25,26] and as a working material for spintronic applications [27]. These observations benefited from the very low magnetic damping in Py [28,29].

The experimental configuration is shown in Fig. 1(a). Ultrashort light pulses were delivered from a Yb-fiber laser operating at $\lambda$=1030 nm fundamental wavelength and at a repetition rate of 50 kHz. Optical excitation was performed with two linearly polarized 300 fs pulses having a wavelength of λ=1030 nm and spot size of 250 μm, with incidence angles of 2° and 60° for pump 1 and pump 2, respectively. Linearly polarized optical probe pulses with a wavelength of λ=515 nm and spot size of 100 μm were obtained by frequency doubling of the fundamental wavelength of the laser. The probe beam was at an angle of incidence of about 5° to the surface normal. The magnetization dynamics was studied by analyzing the Kerr rotation of the polarization plane of the reflected probe beam using a balanced detection scheme. In this geometry, most of the magnetic signal comes from the polar magneto-optical Kerr effect (MOKE), which is sensitive to the out-of-plane component of the magnetization $M_Z$. An external magnetic field of $H$=1 kOe was applied parallel to the sample surface.

An example of the magnetization dynamics excited only with pump 1 and pump 2 is given in Fig.1(b), which exhibits the oscillatory behavior of the $M_Z$ component caused by the magnetization precession. We detect precession dynamics up to 920 ps after the excitation, which is the longest available delay range in our experimental setup. Taking into account different absorption for different incidence angles and hence different incident fluence of $F$= 0.92 mJ/cm$^2$ for pump 1 and $F$=1.5 mJ/cm$^2$ for pump 2 there is no difference in the dynamics excited by a P-polarized or S-polarized pump 1 and for P-polarized pump 2, as expected for a thermal excitation mechanism. Time-resolved traces were fit with a damped sinusoidal function,

$$\Delta M(t) = A \cdot e^{-\frac{t}{\delta}} \cdot sin(2\pi f t - \varphi) , \qquad (1)$$

where $A$ is the amplitude of oscillations, $f$ is the frequency, and $\delta$ is the decay time. The amplitude and frequency of the magnetic oscillations change monotonically with increasing fluence [see Fig.1(c) and (d)] while the decay time remains constant [Fig.1(e)]

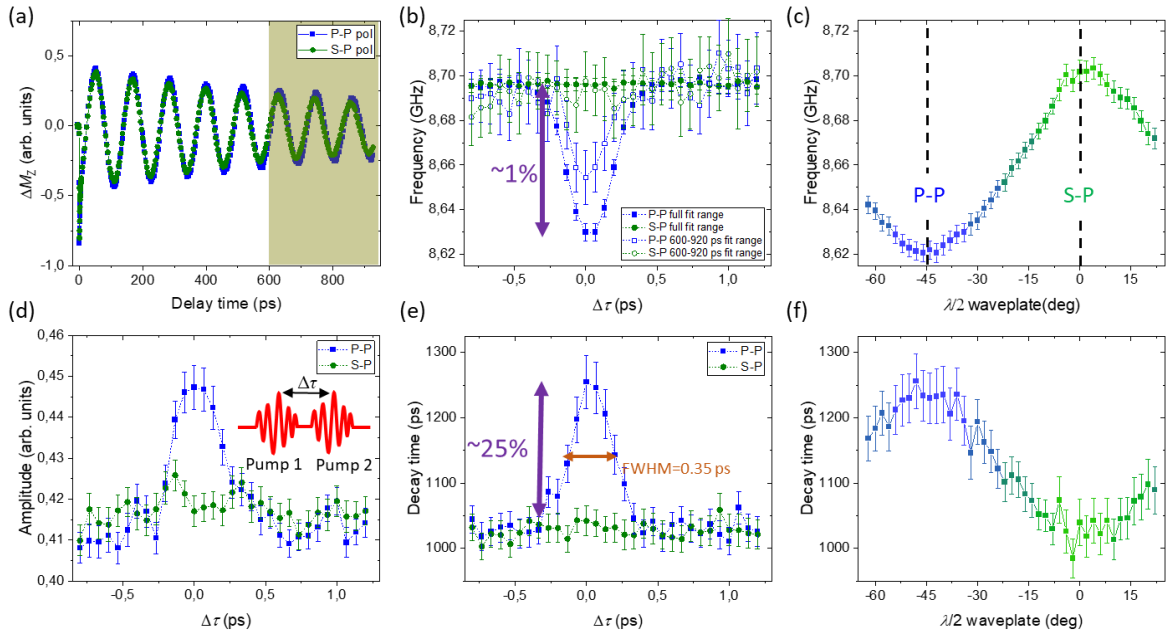

Figure 2. (a) Time-resolved magnetization dynamics after dual-pulse excitation with different pump 1 polarization and with fluencies $F$=1.5 mJ/cm$^2$ for both pump beams. (b), (d) and (e) Frequency, amplitude and decay time of the magnetization precession mode, respectively, as a function of the delay Δτ between two pump pulses with different polarization configurations. (c) frequency and (f) decay time of the magnetization precession as a function of pump 1 polarization direction, determined by the angle of the half-wave plate.

The excitation of the sample with two optical pulses of the same fluence, $F$=1.5 mJ/cm$^2$, leads to different magnetization dynamics. While the dynamics closely resemble those induced by a single optical pulse when pump 1 is S-polarized and pump 2 is P-polarized, a notable divergence occurs when both pump pulses are P-polarized, resulting in distinct oscillations. The frequency, amplitude, and decay time of these oscillations (cf. eq. (1)) as a function of the delay time between the two pump pulses are shown in Fig. 2(b), (d), and (e), respectively. Each panel delineates the dependencies for both the P-P configuration (where both pump pulses have P polarization) and the S-P configuration (where pump 1 is S-polarized, and pump 2 is P-polarized). Remarkably, all three parameters change when the two pump beams temporally overlap in P-P configuration while no changes are observed when the polarizations are orthogonal. The decay time of the precession, when excited with P-polarized pump beams, exhibits an approximately 25% increase compared to the configuration with orthogonal linear polarization and the frequency undergoes a decrease of about 1%. While a change in oscillation frequency could be anticipated due to the different degrees of demagnetization (see Fig. 1(c)), the observed frequency change surpasses expectations based solely on the observed increase in the amplitude of the magnetization precession in the P-P configuration in Fig. 2(d). Consequently, we attribute this observed frequency change to an effect associated with the non-collinear dual optical excitation, specifically linked to interference between the two pump beams. In Fig. 2(c) and (f), we present panels illustrating the frequency and decay time as a function of the polarization direction of pump 1 for $\Delta\tau$=0 ps. Both dependencies follow the superposition of $E$-fields vector projections of the two pumps.

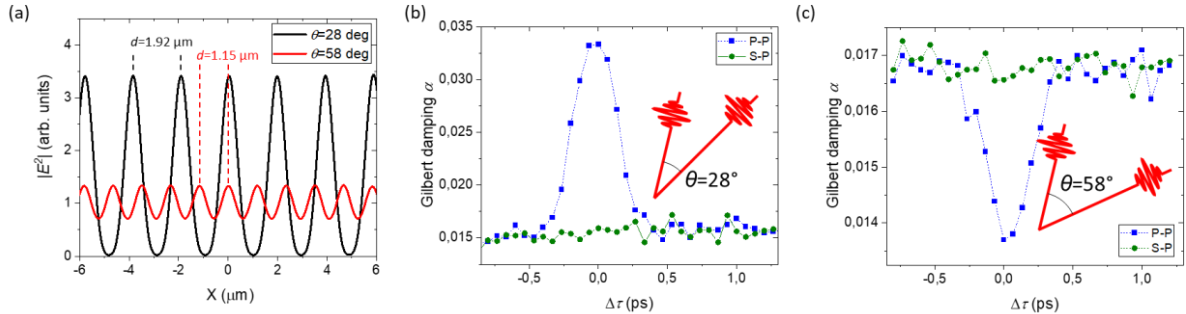

*Figure 3. (a) Calculated spatial distribution of electric field square |E$^2$| when two optical pulses impinge the sample with different separation angle θ. (b) and (c) Gilbert damping parameter α as a function of delay between two pump pulses Δτ for configuration with θ=28 deg and θ=58 deg respectively.*

Two non-collinear optical pump beams, overlapping in time, can induce transient grating through the interference between the two pulses [15, 16, 17]. The spatial modulation of the excitation intensity in a transient grating can potentially alter the behavior of magnetization precession. To evaluate the impact of a transient grating, we varied the angle $\theta$ between the pump pulses, thereby altering the period of the transient grating. The calculated spatial distribution of the electric field square with finite-difference time-domain method using parameters similar to those described in the experimental conditions and for different separation angles is depicted in Fig. 3(a). With $\theta$=28°, a well-defined interference pattern with clear maxima and minima and a period $d$=1.92 μm is formed. In contrast, for $\theta$=58°, the interference pattern is less pronounced, with $d$=1.15 μm close to the wavelength of the pump beams, suggesting a significantly smaller effect of the transient grating for $\theta$=58°.

Experimental data for magnetization dynamics for the $\theta$=28° configuration are presented in the supplementary information, and for $\theta$=58°, they are given in the main text above. We discuss

Gilbert damping parameter $\alpha = 1/(2\pi f \delta)$ as it contains both frequency and decay time when comparing two $\theta$ geometries. Data from SFig. 3(c), (d), and Fig. 2(b), (e) were used to calculate $\alpha$ as a function of $\Delta\tau$ for different $\theta$, as shown in Fig. 3(b) and (c). For both $\theta$ configurations, there is a clear but opposite change in $\alpha$ for P-P pump polarizations when the beams overlap in time. In both cases, the change in $\alpha$ is primarily caused by alterations in the decay time (see SFig. 3(c) and Fig. 2(e)), despite the concurrent decrease in frequency by about 1% observed for both $\theta$ configurations (see SFig. 3(d) and Fig. 2(b)). The spatial modulation of excitation leads to magnetic oscillations with slightly different frequencies on interference maxima and minima, given that the precession frequency is a function of pump power (see Fig. 1(d)). The dephasing between oscillations at maxima and minima during the recovery dynamics likely contributes to the suppression of magnetization precession for $\theta=28°$. Following these arguments, for the $\theta=58°$ configuration, one would expect a smaller increase in $\alpha$ as the difference between maxima and minima is much smaller. Surprisingly, a decrease in $\alpha$ is observed for $\theta=58°$, suggesting that the observed increase in decay time and decrease in frequency of magnetization precession result from interfering optical pulses but not from the transient grating. Furthermore, when performing the fit procedure in the delay range of 600 ps – 920 ps (refer to the shaded area in Fig. 2(a)), where the transient grating should have already dissipated [30], a change in oscillation frequency is still detected (Fig. 2(b) open circles).

The observed changes in frequency and decay time suggest that optical excitation with two pump pulses changes the magnetic configuration of the material. We try to explain the unusual coupling of the dual pump pulses to the magnetization dynamics considering a recently derived relativistic light-spin interaction term in the Hamiltonian [31]

$$\mathcal{H} = \frac{e^2 \hbar}{8m^2 c^2 \omega} \boldsymbol{\sigma} \cdot \text{Re}[-i(\boldsymbol{E} \times \boldsymbol{E}^*)] = -g\mu_B \boldsymbol{\sigma} \cdot \boldsymbol{B}_{opt}, \quad (2)$$

where $\boldsymbol{\sigma}$ denotes the Pauli spin matrixes, $\boldsymbol{E}$ is the total electric field, and $\boldsymbol{B}_{opt}$ the optomagnetic field caused by the electric fields with frequency $\omega$. The total electric field results from the two pump pulses, i.e., $\boldsymbol{E} = \boldsymbol{E}_{pu1} + \boldsymbol{E}_{pu2}$. For the pump pulses we assume plane waves modulated with a Gaussian envelope. This gives for the optomagnetic field (see Supplemental Materials)

$$\boldsymbol{B}_{opt} = \frac{e^2 \hbar}{8m^2 c^2 \omega g \mu_B} \exp\left[-\frac{t^2}{2\Gamma^2} - \frac{(t-\Delta\tau)^2}{2\gamma^2}\right] \text{Re}\left[i\left(\boldsymbol{E}^0_{pu1} \times \boldsymbol{E}^{0*}_{pu2} e^{i(\boldsymbol{k}\cdot\boldsymbol{r}-\omega\Delta\tau)}\right.\right.$$
$$\left.\left. - \boldsymbol{E}^{0*}_{pu1} \times \boldsymbol{E}^0_{pu2} e^{-i(\boldsymbol{k}\cdot\boldsymbol{r}-\omega\Delta\tau)}\right)\right]. \quad (3)$$

The width of the pulses is determined by $\gamma$ and $\Gamma$, $\boldsymbol{k} = \boldsymbol{k}_{pu1} - \boldsymbol{k}_{pu2}$ is the resultant wave vector which is nonzero for noncollinear beams, and $\boldsymbol{E}^0_{pu1}, \boldsymbol{E}^0_{pu2}$ are the electric field vectors of the pulses. It can be recognized from Eq. (3) that the optomagnetic field is maximal when the two pulses fully overlap, i.e., $\Delta\tau = 0$, consistent with the measurements shown above.

Next, we analyze the dependence of the optomagnetic field on the pump polarizations. Starting with the P-P configuration, we compute a nonzero $\boldsymbol{B}_{opt}$ aligned dominantly along the *y* direction (see Supplementary Materials for details). The non-collinearity of the beams is important for obtaining a nonzero $\boldsymbol{B}_{opt}$. The torque due to the optomagnetic field, $\boldsymbol{T} = \boldsymbol{M} \times \boldsymbol{B}_{opt}$, is dominantly in the out-of-plane *z* direction. An increase of the amplitude of the magnetization precession could be caused by the additional torque in the *z* direction. The S-P configuration leads to a nonzero $\boldsymbol{B}_{opt}$ along the *z* direction and a torque $\boldsymbol{T}$ that is along the in-plane *y* direction. This direction can however not lead to a polar MOKE signal, as this is sensitive only to out-of-plane magnetization excursions. The observed dependencies on the pump polarizations are thus consistent with those expected from the light-spin interaction (2).

The light-spin interaction (2) can be seen as a transfer of spin angular momentum from photons to electrons. This additional moment would change the spin-orbit coupling and effectively the magnetic moment of the system, leading to altering the behavior of the magnetic oscillation mode. However, the long-liveness of the induced effect is surprising. Our analysis of the dynamics lasting almost 1 ns indicates that the induced effect is long-lived compared to the 0.35 ps $\Delta\tau$ window where the two pump pulses overlap and which is the acting time of the photomagnetic field $\boldsymbol{B}_{opt}$. It remains currently an open question why the light-induced modification in the electron system after non-collinear optical excitation does not drain to the lattice on a (sub)-picosecond timescale, as expected based on the current knowledge about electron-spin, electron-electron and electron-phonon interaction times [1].

Several prior theoretical investigations have aimed to reveal the interaction of light with spins on the timescale of optical excitation, including predictions of the coupling of the angular moment of light to the spin [32,33]. A similar coupling as in Eq. (2) between the electromagnetic angular moment density and magnetic moments in a solid was phenomenologically proposed [34]. There is experimental evidence of relativistic spin-photon coupling reported in Ref. [32], which was later supported by theoretical investigations [33,35] albeit on the basis of different relativistic interactions. The study of Bigot *et al.* [32] thus provided evidence of an optomagnetic field acting on the electron during pump excitation. However, a single laser pump was used and no long-lasting effects were reported.

In conclusion, our study has demonstrated that using two non-collinearly propagating optical pulses that interfere can elicit enduring changes in the electronic properties of the material, manifesting as alterations in the decay time and frequency of magnetization precession. The observed effect is notably contingent on the relative polarization of the two pulses and is maximized when the excitation pulses overlap in time. The results presented here defy a complete explanation within the current understanding of light-matter interactions. Theoretically, we explore the concept of relativistic light-spin interaction, conceptualized as an optomagnetic field, acting on the timescale of the pump beam duration. This discussion is particularly pertinent in configurations where deviations from the anticipated trends in magnetization dynamics are observed. However, the persistent modification of the magnetic behavior of electrons over an extended duration is unexpected. Hence, we propose the existence of a novel physical effect, wherein the key feature is likely the optical excitation of the medium with interfering light. This phenomenon induces enduring modifications in the magnetic properties of electrons. This work not only introduces new possibilities for the control and manipulation of magnetic states in materials but also suggests potential applications for manipulating electron spin in unforeseen ways. Additionally, our findings offer insights into fundamental interactions between light and matter, with implications that may extend beyond the realm of magnetism. In particular, the significance of considering optical excitation with interfering light for studying and controlling material properties is underscored by our results. Our results encourage further investigations to unravel the underlying mechanisms of this effect and explore its potential applications across diverse material systems.

**Acknowledgments**:

We gratefully thank A. Weber for the assistance in sample preparation. We thank M. Weißenhofer for the valuable discussions. D.P. acknowledges support from the Swiss National

Science Foundation (Project No. 200020_200332). Financial support of the Faculty Research Scheme from IIT (ISM) Dhanbad is acknowledged. This work has been supported by the Swedish Research Council (VR, Grant No. 2022-06725) and the K. and A. Wallenberg Foundation (Grant No. 2022.0079).